\documentstyle[manuscript,aps,epsf]{revtex}
\begin{document}
\preprint{UCONN-97/10\,,\quad NSF-ITP-97-054}

\title{Self-Isospectral Periodic Potentials and Supersymmetric Quantum Mechanics}

\author{Gerald Dunne\footnote{dunne@hep.phys.uconn.edu}}
\address{Physics Department, University of Connecticut, Storrs, CT 06269}
\author{Joshua Feinberg\footnote{joshua@itp.ucsb.edu}}
\address{Institute for Theoretical Physics, University of California, Santa Barbara, CA 93106}
%\date{\today}
\maketitle

\begin{abstract}
We discuss supersymmetric quantum mechanical models with periodic potentials. The important new feature is that it is possible for {\it both} isospectral potentials to support zero modes, in contrast to the standard nonperiodic case where either one or neither (but not both) of the isospectral pair has a zero mode. Thus it is possible to have supersymmetry unbroken and yet also have a vanishing Witten index. We present some explicit exactly soluble examples for which the isospectral potentials have identical band spectra, and which are ``self-isospectral'' in the sense that the potentials have identical shape, but are translated by one half period relative to one another.
\end{abstract}

\vskip 1cm

Supersymmetry and supersymmetry breaking are fundamental issues in theoretical particle physics, and supersymmetric (SUSY) quantum mechanics provides an important testing ground for both physical and computational aspects of SUSY theories \cite{witten,cooper}. There are also many applications to the theory of solitons \cite{solitons}. Of particular interest for particle physics are possible mechanisms for breaking SUSY dynamically. Typically, one considers models with discrete spectra, and then the Witten index, which characterizes the difference between the number of bosonic and fermionic zero modes, may be used to indicate whether or not SUSY is broken \cite{witten}. Interesting subtleties arise for potentials with continuum states \cite{akhoury} or with singularities \cite{jevicki}. 

In this paper we consider SUSY quantum mechanics for {\it periodic} potentials (which therefore have band spectra). The main new feature is that it is possible for the periodic isospectral bosonic and fermionic potentials to have {\it exactly} the same spectrum, {\it including zero modes}. This is in contrast to the usual (nonperiodic and fast decaying) case for which at most one potential of an isospectral pair can have a zero mode.  

Consider one dimensional SUSY quantum mechanical models on the real line. The bosonic and fermionic Hamiltonians $H_\pm$ correspond to an isospectral pair of potentials $V_\pm(x)$ defined in terms of the ``superpotential'' $W(x)$ as
\begin{equation}
V_\pm(x)=W^2(x)\pm W^\prime(x)
\label{iso}
\end{equation}
The Hamiltonians may be factorized into products of hermitean conjugate operators as
\begin{equation}
H_+ = [{d\over dx} + W(x)][-{d\over dx}+W(x)]\,,\quad H_- = [-{d\over dx} + W(x)][{d\over dx}+W(x)]
\label{factorized}
\end{equation}
which indicates that $H_\pm$ are formally positive operators. Thus, their energy spectrum cannot go below zero.\footnote{In some cases where $V_\pm(x)$ are singular $H_\pm$ may have negative energy states \cite{jevicki}. We shall exclude such cases from the present discussion.} The factorization
(\ref{factorized}) also implies that $V_\pm$ have (almost) the same spectrum because there is a one-to-one mapping between the energy eigenstates $\psi_E^{(\pm)}$:
\begin{eqnarray}
\psi_E^{(+)}={1\over \sqrt{E}}\left({d\over dx}+W(x)\right) \psi_E^{(-)};\qquad
\psi_E^{(-)}={1\over \sqrt{E}}\left(-{d\over dx}+W(x)\right) \psi_E^{(+)}
\label{mapping}
\end{eqnarray}
The caveat `almost' is needed above because this mapping between states does 
not apply to the ``zero modes'' (eigenstates with $E=0$), which due to the 
positivity of $H_\pm$, are the lowest possible states in the spectrum. From (\ref{factorized}) it is easy to see that the Schr\"odinger equation $[-\partial_x^2 +V_\pm(x)]\psi_E^{(\pm)}=E\psi_E^{(\pm)}$ has zero modes 
\begin{equation}
\psi_0^{(\pm)}(x)=e^{\pm\int^x W}
\label{zero}
\end{equation}
provided these functions $\psi_0^{(\pm)}$ belong to the Hilbert space. SUSY is said to be unbroken if at least one of the $\psi_0^{(\pm)}$ is a true zero mode. Otherwise, SUSY is said to be broken dynamically. In the broken SUSY case there are no zero modes and so the spectra of $V_\pm$ are identical [due to the mapping (\ref{mapping})].

In the ``standard cases'' \cite{witten}, in which $V_\pm(x)$ tend to positive asymptotic values as $x\rightarrow\pm\infty$, this means that $\psi_0^{(\pm)}$  must be normalizable in order to be true zero modes. But it is clear that in these cases, at most only one of the functions $\psi_0^{(\pm)}$ may be normalizable. Thus the spectra of the potentials $V_\pm$ coincide except possibly for the zero energy ground state level. For example, if the (well-behaved) superpotential tends to asymptotic values with opposite signs as $x\rightarrow\pm\infty$, then one of the zero modes in (\ref{zero}) is normalizable, and SUSY is unbroken; superpotentials that are odd functions, $W(-x)=-W(x)$, belong to this class. Dynamically broken SUSY occurs when the superpotential tends to asymptotic values with equal signs as $x\rightarrow\pm\infty$ ; superpotentials that are even in $x$ are of this type.

Two simple representative examples of unbroken SUSY are: (i) $W(x)=x$, which gives the harmonic oscillator. Of the two functions $\psi_0^{(\pm)}=e^{\pm x^2/2}$, clearly only $\psi_0^{(-)}$ is normalizable and
hence a zero mode. (ii) $W(x)=j\,\tanh x$ (with $j$ a positive integer), which gives the P\"oschl-Teller potentials
\begin{equation}
V_\pm=j^2-j(j\mp 1)sech^2 x
\label{bound}
\end{equation}
Once again, of the two possibilities $\psi_0^{(\pm)}=[\cosh x]^{\pm j}$, only $\psi_0^{(-)}$ is normalizable. $V_-$ has $j$ discrete bound states [with energies $E_n=n(2j-n)$ for $n=0,1,\dots,j-1$] and a continuum beginning at $E=j^2$; on the other hand, $V_+$ has $j-1$ discrete bound states [with 
energies $E_n=n(2j-n)$ for $n=1,\dots,j-1$] and a continuum beginning at $E=j^2$. The two spectra coincide manifestly, except for the zero mode.

 Now consider the superpotential $W(x)$ to be periodic, with period $L$: $W(x+L)=W(x)$. The potentials $V_\pm(x)$ in (\ref{iso}) are therefore also periodic with period $L$. From the Bloch-Floquet theory \cite{kittel}, the Hilbert space consists of quasi-periodic functions: functions that 
satisfy $\psi_k(x+L)=({\rm exp}~ikL)\psi_k(x)$, where the real quantity $k$ is the crystal momentum. 

From (\ref{zero}) we have $\psi_0^{(\pm)}(x+L)=e^{\pm\phi_L} \psi_0^{(\pm)}(x)$ where the real constant $\phi_L$ is given by 
\begin{equation}
\phi_L = \int_x^{x+L} W(y) dy 
\label{phil}
\end{equation}
For either one of the functions $\psi_0^{(\pm)}$ to belong to the Hilbert 
space, we must identify $\pm\phi_L=ikL$. But $\phi_L$ is real, which means that $\phi_L=kL=0$. Thus, the two functions $\psi_0^{(\pm)}$ either {\it both} belong to the Hilbert space, in which case they are strictly periodic with period $L$: $\psi_0^{(\pm)}(x+L)=\psi_0^{(\pm)}(x)$, or (when $\phi_L\neq 0$) {\em neither of them} belongs to the Hilbert space.\footnote{This conclusion is valid provided the $\psi_0^{(\pm)}$ do not have nodes (i.e. zeros, which means that the other function has poles). This condition is violated for example in the case of the Scarf potential \cite{scarf} $V(x)\sim cosec^2(x)$, for which $W(x)=j\,cot x$. We shall exclude such singular potentials from the present discussion, but note that these deserve further study - even in the nonperiodic case, singular superpotentials naturally exhibit interesting properties \cite{jevicki}.} (Note that this is the exact opposite of the situation for nonperiodic potentials where if $\psi_0^{(\pm)}$ is a zero mode of $V_\pm$, then $\psi_0^{(\mp)}$ is {\it not} a zero mode of $V_\mp$.) Thus, in the periodic case the spectra of $V_+$ and $V_-$ match {\it completely}. 

To summarize, we see that
\begin{equation}
\phi_L = \int_0^{L} W(y) dy = 0
\label{necessary}
\end{equation} 
is a necessary condition for unbroken SUSY, and when this condition is
satisfied, the bosonic and fermionic sectors have identical spectra, including
zero modes. The Witten index then vanishes.

It is instructive to consider some simple special classes of periodic superpotentials which satisfy (\ref{necessary}). First, suppose the superpotential is antisymmetric on a half-period:
\begin{equation}
W(x+{L\over 2}) = - W(x)\,,
\label{special1}
\end{equation}
Then, from (\ref{iso}) and (\ref{special1}) we obtain
\begin{equation}
V_\pm(x+{L\over 2}) = V_\mp(x)
\label{shift}
\end{equation} 
The potentials $V_\pm$ are simply translations of one another by half a period, and thus are essentially identical in shape. Therefore, they must support exactly the same spectrum, as SUSY indeed tells us they do. We refer to such a pair 
of isospectral $V_\pm$ that are identical in shape as ``self-isospectral''. A simple example of a superpotential of this type is $W(x)={\rm sin} x$, with
$V_+(x) = {\rm sin}^2 (x) + {\rm cos} x = V_-(x+\pi)$.

Second, consider periodic superpotentials that are even functions of $x$: 
\begin{equation}
W(-x) =  W(x)
\label{special2}
\end{equation}
but which also satisfy the condition $\phi_L=0$ (by subtracting an appropriate constant, any even $W(x)$ can be brought into this class).  The function $dW(x)/dx$ is odd and so (\ref{iso}) implies
\begin{equation}
V_\pm(-x) = V_\mp(x)
\label{reflect}
\end{equation} 
The two potentials are then simply reflections of one another. They have the same shape and therefore give rise to exactly the same spectrum, as we
know from SUSY. Such potentials are also ``self-isospectral''. A simple example of a superpotential of this type is $W(x)={\rm cos} x$, with
$V_+(x) = {\rm cos}^2 (x) - {\rm sin} x = V_-(-x)$.

Third and last, consider periodic superpotentials that are odd functions of $x$: 
\begin{equation}
W(-x) = - W(x)
\label{special3}
\end{equation}
Then $\phi_L=0$ is satisfied trivially. The function $dW(x)/dx$ is even and thus $V_\pm(x)$ are also even. In this case, $V_\pm(x)$ are not necessarily related by simple translations or reflections. They are 
isospectral, but may not be ``self-isospectral". As an example, the
superpotential $W(x)= {\rm sin}x + {\rm sin} 2x $ gives rise to an isospectral pair which is not self-isospectral, while $W(x)= {\rm sin}x + {\rm sin} 3x $
[which also belongs to the first special class mentioned above: $W(x+\pi)=-W(x)$] gives rise to a self-isospectral pair.

To make these general ideas more explicit, we now present a class of exactly soluble models. We illustrate this class beginning with the simplest case. Consider the superpotential
\begin{equation}
W(x)=m{sn(x|m)cn(x|m)\over dn(x|m)}
\label{super}
\end{equation}
Here $sn(x|m)$, $cn(x|m)$ and $dn(x|m)$ are the Jacobi elliptic functions \cite{abram,whittaker}, and the (real) elliptic modulus parameter $m$ can be chosen $0< m\leq 1$. Given this superpotential, the isospectral pair (\ref{iso}) of potentials is
\begin{equation}
V_\pm=\left\{\matrix{2-m+2(m-1)/ dn^2(x|m)\cr 2-m-2 dn^2(x|m)}\right.
\label{isopots}
\end{equation}
Some relevant properties of the Jacobi elliptic functions are listed here. 

1. Periodicity properties:
\begin{eqnarray}
sn(x+2 K(m)|m)&=&-sn(x|m)\nonumber\\
cn(x+2 K(m)|m)&=&-cn(x|m)\nonumber\\
dn(x+2 K(m)|m)&=&dn(x|m)
\label{periodicity}
\end{eqnarray}
Here $K(m)$ is the ``real elliptic quarter period'': $K(m)\equiv\int_0^{\pi/2}d\theta/\sqrt{1-m\sin^2\theta}$.

2. Differentiation properties:
\begin{eqnarray}
{d\over dx} sn(x|m)&=& cn(x|m)\, dn(x|m)\nonumber\\
{d\over dx} cn(x|m)&=& -sn(x|m)\, dn(x|m)\nonumber\\
{d\over dx} dn(x|m)&=& -m\, sn(x|m)\, cn(x|m)
\label{diff}
\end{eqnarray}

3. Quadratic relations:
\begin{equation}
-dn^2(x|m)+1-m=-m\, cn^2(x|m)=m\, sn^2(x|m)-m
\label{quadratic}
\end{equation}

Finally, we note that when $m=1$ these relations all reduce to those for the familiar hyperbolic functions since
\begin{equation}
sn(x|1)= tanh x ;\qquad cn(x|1)=sech x ;\qquad dn(x|1)=sech x
\label{hyperbolic}
\end{equation}
Thus, when $m=1$ the superpotential in (\ref{super}) reduces to $tanhx$ and the isospectral potentials (\ref{isopots}) reduce to the $j=1$ case of the example in (\ref{bound}). [The reader is urged to consider the $m\to 1$ limit at all stages of the subsequent discussion.]

From the periodicity properties (\ref{periodicity}), the superpotential $W(x)$ in (\ref{super}) and the potentials $V_\pm$ in (\ref{isopots}) have period $2K(m)$:
\begin{equation}
V_\pm (x+2K(m))=V_\pm(x)
\label{period}
\end{equation}
The zero modes (\ref{zero}) are \footnote{Note that the function $dn(x|m)$ has no nodes or poles on the real axis.}
\begin{equation}
\psi_0^{(\pm)}(x)=e^{\mp log\, dn(x|m)} = [dn(x|m)]^{\mp 1}
\label{isozero}
\end{equation}
Both $\psi_0^{(+)}$ and $\psi_0^{(-)}$ have period $2K(m)$, and are {\it both} good zero modes. Thus the spectra of $V_\pm$ should be {\it identical}. This 
can be checked explicitly because the spectrum can be computed exactly, since  
the Schr\"odinger equation for the potentials (\ref{isopots}) is an example of the Lam\'e equation, whose explicit solution is known in terms of elliptic functions \cite{whittaker}. Each spectrum has a single bound band and a continuum, as shown in Fig. 1.

The upper edge of the bound band has energy $E_1=1-m$, with Bloch wavefunctions
\begin{equation}
\psi_1^{(\pm)}=\left\{\matrix{sn(x|m)/dn(x|m)\cr cn(x|m)}\right.
\label{upper}
\end{equation}
while the the lower edge of the continuum band has energy $E_2=1$, and Bloch wavefunctions
\begin{equation}
\psi_2^{(\pm)}=\left\{\matrix{cn(x|m)/dn(x|m)\cr sn(x|m)}\right.
\label{lower}
\end{equation}
These band-edge properties may be verified directly using the various properties listed in (\ref{diff},\ref{quadratic}).

Concentrating on the potential $V_-$ [an analogous analysis holds for $V_+$], the Schr\"odinger equation can be written [using (\ref{quadratic})] in Lam\'e form
\begin{equation}
\psi^{\prime\prime}=[2m\,sn^2(x|m)-m-E]\psi
\label{lame}
\end{equation} 
This equation has two independent solutions
\begin{equation}
\psi(x)={H(x\pm \alpha)\over \Theta(x)}e^{\mp x Z(\alpha)}
\label{exact}
\end{equation}
where the parameter $\alpha$ is related to the energy eigenvalue $E$ by $E=dn^2(\alpha|m)$; $H(x)$ is the Jacobi eta function, $\Theta(x)$ the Jacobi theta function, and $Z(\alpha)$ the Jacobi zeta function~\cite{whittaker}. It is an instructive exercise to verify that at the band edges these solutions reduce to the wavefunctions in (\ref{isozero},\ref{upper},\ref{lower}), and furthermore that when $m=1$ they reduce to the well-known bound state and continuum states for the P\"oschl-Teller potential $V_-=1-2 tanh^2x$.

Given the exact solution (\ref{exact}) we can use Bloch's theorem to find the exact dispersion relation between the energy $E$ and the crystal momentum $k$:
\begin{eqnarray}
\psi(x+2K(m))&=&e^{i k 2K(m)} \psi(x)\nonumber\\
k&=&\mp{\pi\over 2K(m)}\pm iZ(dn^{-1}(\sqrt{E}|m))
\label{crystal}
\end{eqnarray}
We plotted this dispersion relation in Fig. 2, which clearly shows the band-gap. Note that it is rare to have an exact solution for these band features.

The isospectral potentials $V_\pm$ in (\ref{isopots}) are also self-isospectral. Indeed, using the properties
\begin{eqnarray}
sn(x+ K(m)|m)&=&cn(x|m)/dn(x|m)\nonumber\\
cn(x+ K(m)|m)&=&-\sqrt{1-m}\,sn(x|m)/dn(x|m)\nonumber\\
dn(x+ K(m)|m)&=&\sqrt{1-m}/dn(x|m)
\label{half}
\end{eqnarray}
we see that the superpotential (\ref{super}) satisfies the condition (\ref{special1}), and the two potentials are identical up to a displacement by half a period (see Fig. \ref{iso3plot}):
\begin{equation}
V_+(x+K(m))=V_-(x)
\label{self}
\end{equation}
Since each potential extends indefinitely and periodically, they are indistinguishable as fas as their spectrum is concerned.

This raises the question of what happens in the $m\to 1$ limit, because we know that when $m=1$ the potentials $V_\pm$ are genuinely different and only $V_-$ has a zero mode. The situation is best illustrated by Fig. 4. Consider a single period $-K(m)\leq x\leq K(m)$. As $m\to 1$, $K(m)\to\infty$, and this single period becomes our real line. On this domain, the potential $V_-$ becomes $1-2tanh^2x$, and its bound band collapses smoothly into a single discrete bound level (see Fig. 1). Moreover, the Bloch wavefunctions (\ref{isozero}) and (\ref{upper}) at the lower and upper edges of the bound band each tend smoothly to the normalizable wavefunction $\psi_0^{(-)}=sechx$ of this single bound state. On the other hand, on this domain $V_+$ flattens out and becomes $1$, which has no bound states (only a continuum $E>1$). Correspondingly, the Bloch wavefunctions (\ref{isozero}) and (\ref{upper}) at the lower and upper edges of the bound band tend smoothly to $coshx$ and $sinhx$ (respectively), which are {\it not} normalizable. Displacing this picture by $K(m)$ (i.e. by half a period), the roles of $V_-$ and $V_+$ are interchanged.

This is just the simplest example of a general class of exactly solvable periodic potentials with bound bands. Indeed, it is a classic result that the spectrum of the Lam\'e equation
\begin{equation}
\psi^{\prime\prime}=[j(j+1)m\,sn^2(x|m)-E]\psi
\label{genlame}
\end{equation}
has $j$ bound bands and a continuum band \cite{whittaker}. Moreover, the exact solution [analogous to (\ref{exact})] can be written in terms of elliptic functions (although for $j\geq 2$ the relation between the energy and the crystal momentum becomes more difficult to specify explicitly). In order to make the connection between these Lam\'e equations and SUSY quantum mechanics we must shift the Lam\'e potential by a constant to ensure that the lower edge of the lowest band has energy $E=0$. For example, for $j=2$ the self-isospectral pair of potentials is
\begin{equation}
V_\pm=\left\{\matrix{4m-2+2\sqrt{m^2-m+1}+6(m-1)/dn^2(x|m)\cr 4m-2+2\sqrt{m^2-m+1}-6 dn^2(x|m)}\right.
\label{twoband}
\end{equation}
The energy spectrum is shown in Fig. 5, and the band-edge Bloch wavefunctions for $V_-$ are
\begin{eqnarray}
E_0=0&:\quad&\psi_0^{(-)}=m+1+\sqrt{m^2-m+1}-3msn^2(x|m)\nonumber\\
E_1=-1-m+2\sqrt{m^2-m+1}&:\quad&\psi_1^{(-)}=cn(x|m)dn(x|m)\nonumber\\
E_2=-1+2m+2\sqrt{m^2-m+1}&:\quad&\psi_2^{(-)}=sn(x|m)dn(x|m)\nonumber\\
E_3=2-m+2\sqrt{m^2-m+1}&:\quad&\psi_3^{(-)}=sn(x|m)cn(x|m)\nonumber\\
E_4=4\sqrt{m^2-m+1}&:\quad&\psi_4^{(-)}=m+1-\sqrt{m^2-m+1}-3msn^2(x|m)
\label{twospectrum}
\end{eqnarray}

The band-edge Bloch wavefunctions $\psi_n^{(+)}$ for $V_+$ are obtained simply by shifting the $\psi_n^{(-)}$ in (\ref{twospectrum}) by half a period. The superpotential for the isospectral pair (\ref{twoband}) is determined by the zero-mode $\psi_0^{(-)}$:
\begin{equation}
W=-{d\over dx}\log\,\psi_0^{(-)}={6 m sn(x|m)cn(x|m)dn(x|m)\over m+1+\sqrt{m^2-m+1}-3m sn^2(x|m)}
\label{supertwo}
\end{equation}
When $m=1$ this reduces to $W=2 tanhx$, which is the $j=2$ P\"oschl-Teller model in (\ref{bound}).

This procedure may be repeated for higher integer values of $j$ in the Lam\'e equation (\ref{genlame}), leading to a general class of self-isospectral periodic potentials. The band-edge Bloch wavefunctions (there are $2j+1$ of them since there are $j$ bound bands) are always polynomials of order $j$ in the Jacobi elliptic functions (known as Lam\'e functions \cite{whittaker}). Thus, it is a straightforward algebraic problem to determine the band-edge wavefunctions and energies. Indeed, in a beautiful paper \cite{iachello}, Alhassid et al showed that the band-edge energies are simply the eigenvalues of the $su(2)$ operator $J_x^2+mJ_y^2$, with $J_x$ and $J_y$ being the standard $su(2)$ generators in a $(2j+1)$ dimensional matrix reprentation (see also \cite{ward}).

Another generalization of the single bound band example (\ref{isopots}) is obtained by generalizing the superpotential (\ref{super}) to
\begin{equation}
W(x)=j\,m{sn(x|m)cn(x|m)\over dn(x|m)}
\label{jsuper}
\end{equation}
where $j$ is a positive integer. The resulting self-isospectral potentials, $V_\pm=W^2\pm W^\prime$, are 
\begin{equation}
V_\pm=j^2(2-m)-j(j\mp 1)dn^2(x|m)+j(j\pm 1)(m-1)/ dn^2(x|m)
\label{jisopots}
\end{equation}
The zero mode Bloch wavefunctions are $\psi_0^{(\pm)}=[dn(x|m)]^{\mp j}$. The potentials (\ref{jisopots}) are self-isospectral: $V_+(x+K(m))=V_-(x)$, as can be seen from Fig. 6. Note also that as $m\to 1$, in the domain $-K(m)\leq x\leq K(m)$, $V_-$ approaches the P\"oschl-Teller potential $V_-$ in (\ref{bound}) which has $j$ discrete bound states, while $V_+$ approaches the P\"oschl-Teller potential $V_+$ in (\ref{bound}) which has $j-1$ discrete bound states.

We conclude by noting that it would be interesting to extend some of these ideas to field theoretic examples.
\vskip .5in
\noindent{\bf Acknowledgements}
This work has been supported by the DOE grant DE-FG02-92ER40716.00 (GD) and NSF grant PHY89-04035 (JF).

\begin{figure}
    \epsffile{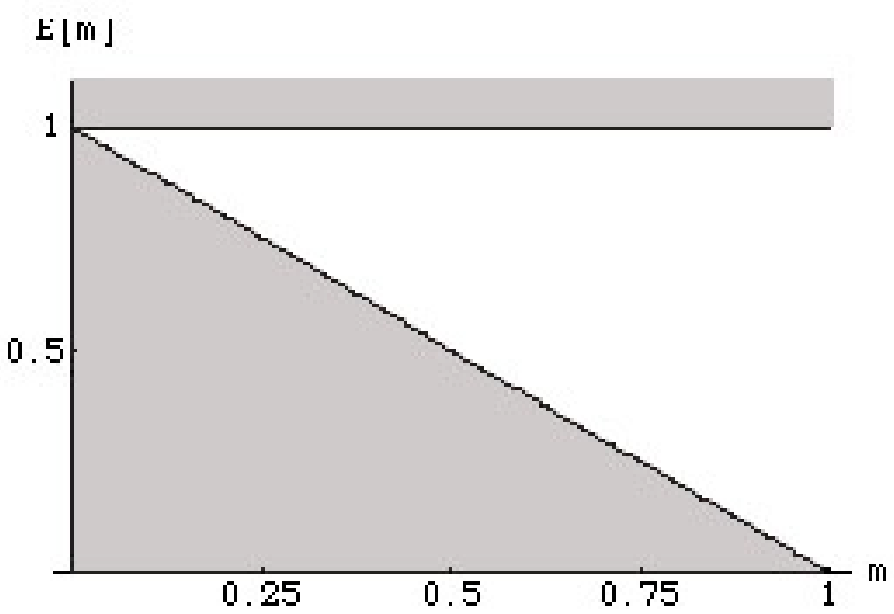}
    \caption{The spectrum of the isospectral pair of potentials in (\protect{\ref{isopots}}), as a function of the elliptic parameter $m$. There is a single bound band, bounded below by energy $E_0=0$, and above by energy $E_1=1-m$. There is also a continuum band beginning at $E_2=1$. Note that when $m=1$ the bound band smoothly degenerates into a single discrete bound level of energy $E_0=0$; this is just the zero mode of the $j=1$ P\"oschl-Teller system in (\protect{\ref{bound}}).}
  \label{iso1plot}
\end{figure}

\begin{figure}
    \epsffile{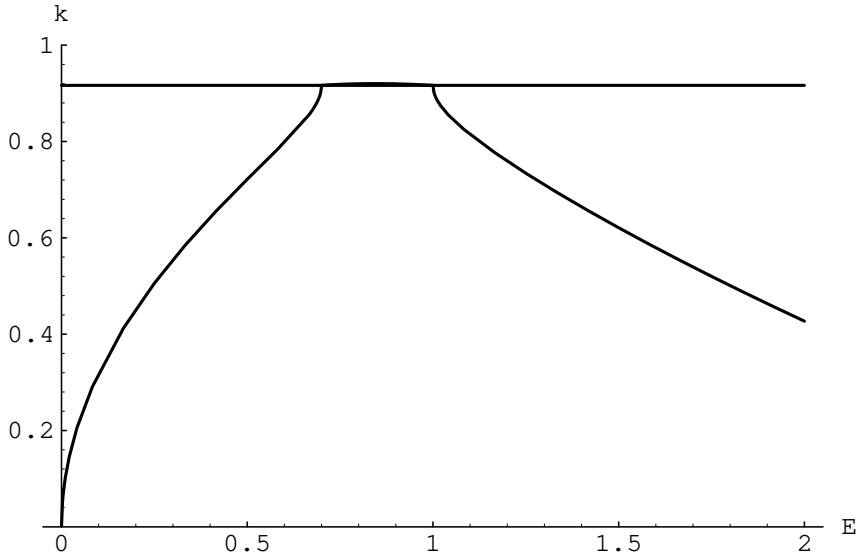}
    \caption{The {\it exact} dispersion relation (\protect{\ref{crystal}}) between energy $E$ and crystal momentum $k$ for the isospectral system (\protect{\ref{isopots}}). This plot is for $m=0.3$, and we clearly see the band gap between $E=0.7$ and $E=1$. The horizontal line marks the edge of the Brillouin zone, at which $k=\pi/(2 K(m))$. }
  \label{iso2plot}
\end{figure}

\begin{figure}
    \epsffile{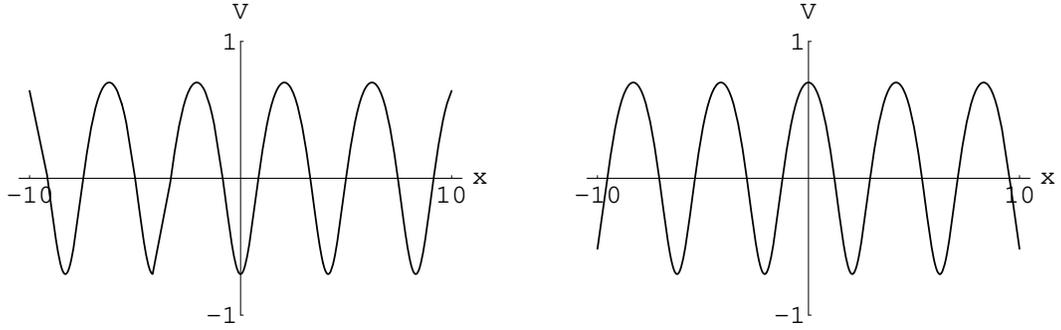}
    \caption{The self-isospectral potentials (\protect{\ref{isopots}}): $V_-$ (left) and $V_+$ (right). Note that they are identical, except for being displaced by half a period. These plots are for elliptic parameter $m=0.7$.}
  \label{iso3plot}
\end{figure}

\begin{figure}
    \epsffile{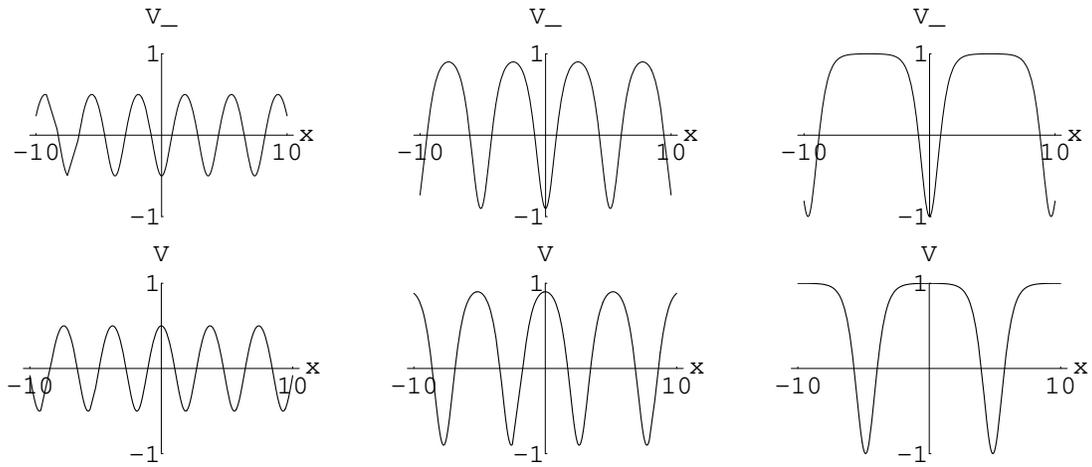}
    \caption{The upper (lower) level shows the potential $V_-$ ($V_+$) for various values of the elliptic modulus parameter $m=0.5$, $m=0.9$, and $m=0.999$, going from left to right. Notice that $V_+$ is identical to $V_-$, but shifted by half a period. Also note that as $m$ approaches $1$, within the central period, $-K(m)\leq x\leq K(m)$, $V_-$ approaches a P\"oschl-Teller potential with a single binding well, while $V_+$ flattens out to a constant. The situation is interchanged if we displace the picture by $K(m)$, which is half a period of the potentials.}
  \label{iso4plot}
\end{figure}

\begin{figure}
    \epsffile{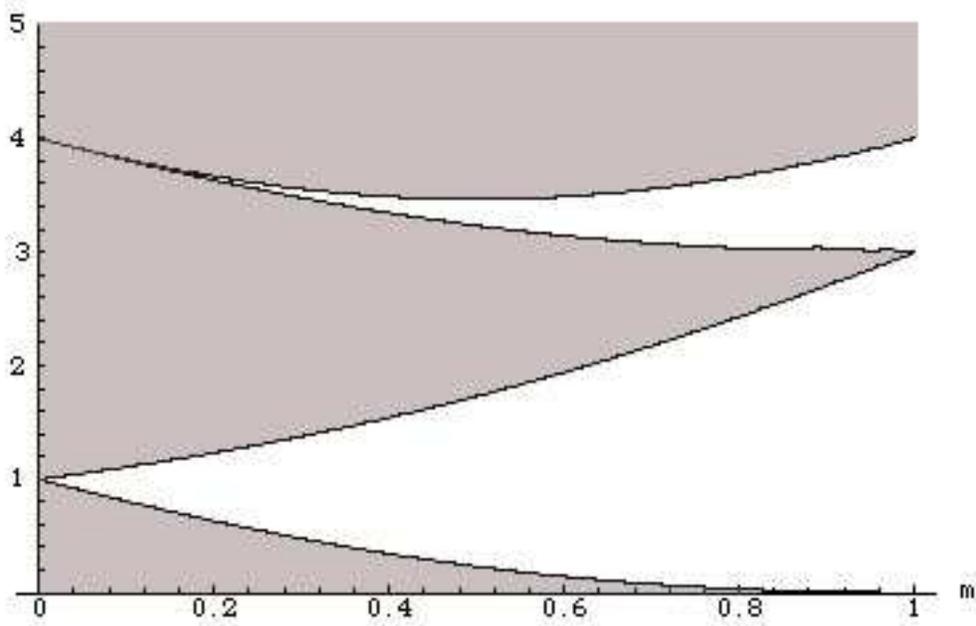}
    \caption{The spectrum of the isospectral pair of potentials in (\protect{\ref{twoband}}), as a function of the elliptic parameter $m$. There are two bound bands, and a continuum band beginning at $E=4\protect{\sqrt{m^2-m+1}}$. Note that when $m=1$ the bound bands smoothly degenerate into two discrete bound levels of energy $E=0$ and $E=3$ and a continuum threshold at $E=4$; these are the two bound states and continuum threshold of the $j=2$ P\"oschl-Teller system in (\protect{\ref{bound}}).}
  \label{iso5plot}
\end{figure}

\pagebreak

\begin{figure}
    \epsffile{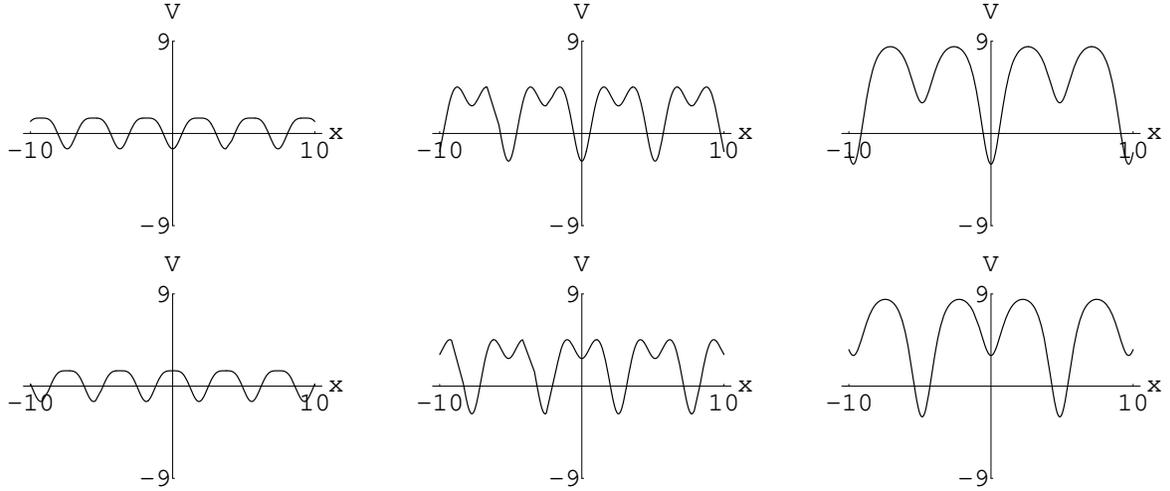}
    \caption{The upper (lower) level shows the potential $V_-$ ($V_+$) in (\protect{\ref{jisopots}}) for $j=3$ and for various values of the elliptic modulus parameter $m=0.5$, $m=0.9$, and $m=0.999$, going from left to right. Notice that $V_+$ is identical to $V_-$, but shifted by half a period. Also note that as $m$ approaches $1$, within the central period, $-K(m)\leq x\leq K(m)$, $V_-$ approaches the $j=3$ P\"oschl-Teller potential, while $V_+$ approaches the $j=2$ P\"oschl-Teller potential. The situation is interchanged if we displace the picture by $K(m)$, which is half a period of the potentials.}
  \label{iso6plot}
\end{figure}

\end{document}